\documentclass[aip,jcp,10pt,twocolumn,superscriptaddress,footinbib]{revtex4-1} \usepackage{graphicx}

\usepackage[latin1]{inputenc}

\usepackage{amsmath}
\usepackage{amssymb}
\usepackage{mathrsfs} 
\usepackage{color}
\usepackage{graphicx,color,colortbl}
\usepackage{rotating,array,tabularx,booktabs}
\usepackage{varwidth,xcolor}

\newcommand{\bvec}[1]{{\bf\string#1 }}
\newcommand{\upd}{\mathrm{d}}

\begin{document}
 
\title{ \LARGE Phase diagram of two-dimensional hard rods from fundamental mixed measure density functional theory} 

\author{Ren\'e Wittmann}
\thanks{These two authors contributed equally}
\affiliation{Department of Physics, University of Fribourg, CH-1700 Fribourg, Switzerland}

\author{Christoph E. Sitta}
\thanks{These two authors contributed equally}
\affiliation{Institut f{\"u}r Theoretische Physik II, Weiche Materie,
Heinrich-Heine-Universit{\"a}t D{\"u}sseldorf, D-40225 D{\"u}sseldorf, Germany}

\author{Frank Smallenburg}
\affiliation{Institut f{\"u}r Theoretische Physik II, Weiche Materie,
Heinrich-Heine-Universit{\"a}t D{\"u}sseldorf, D-40225 D{\"u}sseldorf, Germany}

\author{Hartmut L\"owen}
\affiliation{Institut f{\"u}r Theoretische Physik II, Weiche Materie,
Heinrich-Heine-Universit{\"a}t D{\"u}sseldorf, D-40225 D{\"u}sseldorf, Germany}

\date{\today}
\begin{abstract}
 A density functional theory for the bulk phase diagram of two-dimensional orientable hard rods is proposed and tested against Monte Carlo computer simulation data. In detail, an explicit density functional is derived from fundamental mixed measure theory and freely minimized numerically for hard discorectangles. The phase diagram, which involves stable isotropic, nematic, smectic and crystalline phases, is obtained and shows good agreement with the simulation data. Our functional is valid for a multicomponent mixture of hard particles with arbitrary convex shapes and provides a reliable starting point to explore various inhomogeneous situations of two-dimensional hard rods and their Brownian dynamics.
\end{abstract}

\maketitle

\section{Introduction}

Classical density functional theory (DFT) of inhomogeneous fluids \cite{Evans1979} provides a microscopic theory for freezing, for reviews see \cite{Oxtoby1991, Singh1991, Loewen1994a, WuL2007, TarazonaCMR2008, Roth2010}. This has been exploited for spherical particles with radially symmetric pairwise potentials (such as hard or soft spheres) both in three \cite{BrykRMD2003, KoenigRM2004, TroesterOBVB2012} and two \cite{RothMO2012} spatial dimensions where the freezing line of liquids has been predicted which good accuracy.
Density functional theory of freezing can also be formulated for orientational degrees of freedom as documented in Onsager's seminal work for the isotropic--nematic transition \cite{Onsager1949}. This has been applied to investigate the stability of liquid crystalline phases (such as isotropic, nematic, smectic) in three \cite{PoniewierskiH1988, vanRoijBMF1995, Bohle1996, GrafL1999, CinacchiS2002, HansenGoosM2009, HansenGoosM2010, WittmannMM2014, WittmannMM2015, OettelKDESHG2016} and in two \cite{VargaS1998, MartinezRatonVM2005, WensinkV2007, delasHerasVM2009, VargaGAPQH2009,Chen2013, MuellerdlHRH2015, OettelKDESHG2016, DiazDeArmasMR2017} dimensions. For both translational and orientational degrees of freedom, many different "meso-phases" with partial translational or orientational order are conceivable and therefore the resulting phase diagram is typically much more complex \cite{BolhuisF1997}.

The most elaborate DFTs were derived for hard particles which possess only steric or excluded-volume interactions. In these systems, temperature scales out such that the density (or packing fraction) is the only remaining parameter apart from the  particle shape. In particular, the Fundamental Measure Theory (FMT) originally invented by Rosenfeld \cite{Rosenfeld1989} 
has proven to be very successful for hard-body fluids in three dimensions, including the isotropic phase of particles with non-spherical shape~\cite{Rosenfeld1994, MarechalL2013}. 
The basic input into FMT are different weighted densities, which depend only on the geometry of a single body. 
This simple structure allows for an efficient numerical implementation.
The versatile framework of Rosenfeld's FMT allows to use the same building blocks to construct a new version yielding a more accurate equation of state \cite{HansenGoosR2006} and, upon introducing additional weighted densities, to obtain generalized functionals for freezing \cite{TarazonaR1997,RothMO2012} and liquid crystal phases~\cite{HansenGoosM2009,HansenGoosM2010,WittmannMM2014}.

The usual first step to derive FMT is to take the low-density limit and decompose the Mayer function of the hard-core interaction. However, there is no exact representation based on a finite number of weighted densities in two (and other even) dimensions \cite{Rosenfeld1990},
or in any dimension if the shape of the freely-rotating bodies is anisotropic~\cite{Rosenfeld1994}.
Instead, for two-dimensional hard disks (HD)~\cite{RothMO2012} and arbitrary convex bodies in three dimensions~\cite{HansenGoosM2009} an infinite series of tensorial weighted densities is necessary, which, for practical reasons, is usually truncated after the term including rank-two tensors. A more sophisticated expansion can be defined in terms of orthonormal functions, such as spherical harmonics in three dimensions~\cite{MarechalDD2017}.
 
Regarding the ongoing progress in numerical techniques and computer speed, versions of FMT based on two-body weighted densities \cite{TarazonaR1997,MartinezRatonCC2008,WittmannMM2015,WittmannMM2015b}, which are exact in the low-density limit, become a valid alternative to an approximate treatment, particularly in two dimensions.
Another tractable functional involving many-body measures has been derived for infinitely thin disks in three dimensions \cite{EsztermannRS2006}. 
The most general formulation of FMT for mixtures of arbitrary convex bodies in any dimension is the so-called Fundamental Mixed Measure Theory (FMMT) \cite{WittmannMM2015,WittmannMM2015b}.

In this paper, we consider an explicit DFT based on FMMT for a simple model system of orientable hard rods in two spatial dimensions. 
We study particles with a "discorectangular" shape (the two-dimensional analogue of spherocylinders) whose phase diagram is spanned by their packing fraction and aspect ratio only, while also considering the HD limit.
Monte Carlo (MC) computer simulation \cite{AllenT1989} data are available for the bulk phase diagram of these discorectangles \cite{BatesF2000} and involve an isotropic, nematic and crystalline phase.
Here we evaluate our FMMT functional analytically and numerically and obtain a bulk phase diagram. In doing so, we also extend the previous MC data \cite{BatesF2000} and resolve between a two-dimensional smectic and a full crystalline phase. Our DFT reproduces the topology of this enhanced phase diagram. It is therefore the first functional which gets the stability of four liquid-crystalline phases simultaneously in two dimensions. The FMMT functional is given in a general form for multicomponent mixtures of arbitrary convex hard particles. It can serve as an input for future DFT studies of two-dimensional liquid crystals at interfaces \cite{PatricioRESBTdG2011} and for Brownian dynamics of rods \cite{RexWL2007} similar in spirit as the FMT functional for HD proposed by Roth and coworkers \cite{RothMO2012}.

The paper is organized as follows: in Sec.~\ref{sec:theory} we derive an analytical expression for the DFT functional. Our MC simulations and our numeric DFT minimization are described in Sec.~\ref{sec:methods} and results of our calculations are presented and discussed in Sec.~\ref{sec:results}. We conclude in Sec.\ref{sec:conclusions}.

\section{Density functional theory}
\label{sec:theory}
 
To tackle the general case first, we consider a two-dimensional system of $\kappa$ components of anisotropic particles.
The equilibrium configuration of the particles of each species $i$ is described by a density profile $\rho_i(\mathcal{R})\equiv\rho_i(\bvec{r},\varphi)$ which depends on position $\bvec{r}$ and orientation $\varphi$.
For any external potential $V_i^{\text{ext}}(\mathcal{R})$ acting on the particles, the fundamental variational principle $\delta\Omega/\delta\rho_i(\mathcal{R})=0$ of DFT \cite{Evans1979} states that the unique equilibrium densities minimize the functional
\begin{equation}
 \Omega[\{\rho_i\}]=\mathcal{F}[\{\rho_i\}]+\sum_{i=1}^{\kappa}\int\upd\mathcal{R}\rho_i(\mathcal{R})(V_i^{\text{ext}}(\mathcal{R})-\mu_i)\,,
 \label{eq_functional}
\end{equation}
which then equals the grand potential $\Omega$ of the system. The short notation $\int\!\upd\mathcal{R}$ denotes the integral $\int_{\mathbb{R}^2}\upd\bvec{r}$ over all positions and the orientational average $\int_0^{2\pi}\frac{\upd\varphi}{2\pi}$ and $\mu_i$ denote the chemical potentials of each species. 

The intrinsic free energy
\begin{equation}
\beta\mathcal{F}[\{\rho_i\}]=\beta\mathcal{F}_{\text{id}}+\beta\mathcal{F}_{\text{exc}}=\int\upd\bvec{r}(\Phi_{\text{id}}(\bvec{r})+\Phi_{\text{exc}}(\bvec{r}))
\label{eq_Fintr}
\end{equation}
or its density $\Phi(\bvec{r})$ is usually separated into excess ($\mathcal{F}_{\text{exc}}$) and ideal-gas ($\mathcal{F}_{\text{id}}$) contributions.
The density of the latter reads
$\Phi_\text{id}(\bvec{r})=\sum_{i=1}^{\kappa}\int_0^{2\pi}\frac{\upd\varphi}{2\pi}\rho_i(\bvec{r},\varphi) \left(\ln (\rho_i(\bvec{r},\varphi)\Lambda^2)-1\right)$,
with the thermal wavelength $\Lambda$ and the inverse temperature $\beta^{-1}=k_{\text{B}}T$.

In order to derive the excess free energy density $\Phi_{\text{exc}}$ for a system with hard interactions along the lines of FMT~\cite{Rosenfeld1989} we consider the exact functional
\begin{equation}
 \label{eq_ldlimit}  \beta \mathcal{F}_{\mathrm{exc}} \rightarrow - \frac{1}{2} \sum_{i,j=1}^{\kappa}\iint \upd\mathcal{R}_1\: \upd\mathcal{R}_2 \: \rho_i(\mathcal{R}_1) \:\rho_j(\mathcal{R}_2)\: f_{ij}(\mathcal{R}_1, \mathcal{R}_2)
\end{equation}
in the dilute limit $\rho_i\to 0$, where only the interactions between two particles are relevant.
These are represented by the Mayer function
\begin{equation}
 \label{eq_fijDEF}
 f_{ij}(\mathcal{R}_1, \mathcal{R}_2) =e^{-\beta U_{ij}}-1= \left\{ \begin{array}{cl} 0 & \quad {\rm if} \;\; \mathcal{B}_i \cap \mathcal{B}_j = \emptyset \\
 -1 & \quad {\rm if } \;\; \mathcal{B}_i \cap \mathcal{B}_j \neq \emptyset  
 \end{array}\right.
\end{equation}
of two hard bodies $\mathcal{B}_i$ and $\mathcal{B}_j$ with the pair interaction potential $U_{ij}(\mathcal{R}_1, \mathcal{R}_2)$.
Since this interaction only depends on whether the intersection
\begin{equation}
 \mathcal{I}_{ij}(\mathcal{R}_1, \mathcal{R}_2):=\mathcal{B}_i(\mathcal{R}_1)\cap \mathcal{B}_j(\mathcal{R}_2)
\end{equation}
is the empty set $\emptyset$ or not, Eq.~\eqref{eq_ldlimit} can be simplified using purely geometrical arguments to rewrite $f_{ij}(\mathcal{R}_1, \mathcal{R}_2)$ in terms of quantities that are functions of $\mathcal{R}_1$ or $\mathcal{R}_2$ only.

\subsection{Mayer function of two dimensional hard bodies}
 
For two-dimensional HD mixtures an exact decomposition of the Mayer function from Eq.~\eqref{eq_fijDEF} 
can be found by means of
(i) simple geometrical considerations \cite{MartinezRatonCC2008}, 
(ii) the Gauss-Bonnet theorem from differential geometry \cite{RothMO2012} or
(iii) the translative integral formula \cite{WittmannMM2015,WittmannMM2015b} from integral geometry \cite{SchneiderW2008}.
Considering now mixtures of arbitrary convex bodies in two dimensions, we will show that both strategies (ii) and (iii) 
lead to the same decomposition as for HD, in the sense that all terms are still present in the HD limit. 
Quite in contrast, the hard-sphere limit in three dimensions can be simplified to a deconvolution in terms of one-body weighted densities \cite{HansenGoosM2009,HansenGoosM2010,WittmannMM2015,WittmannMM2015b}. 
The origin of FMMT lies in strategy (iii), since it provides the proper mathematical foundation of employing two-body weighted densities.
There are some alternative ways to derive such a functional   
from (iv) zero dimensional cavities \cite{MarechalGHL2011} or (v) an approximate virial series \cite{MarechalKM2014},
which imply the same decomposition of the Mayer function for anisotropic bodies.

Following Rosenfeld \cite{Rosenfeld1990}, we define the three scalar weight functions
\begin{align} 
 \label{eq_wfscal}
 \omega_i^{(2)}(\mathcal{R}) &= \Theta\left(|\mathbf{R}_i(\hat{\mathcal{R}})|-|\mathbf{r}|\right) \, , \cr
 \omega_i^{(1)}(\mathcal{R}) &= \frac{\delta(|\mathbf{R}_i(\hat{\mathcal{R}})|-|\mathbf{r}|)}{ \bvec{n}_i(\hat{\mathcal{R}})\cdot\hat{\bvec{r}}} \, , \cr  
 \omega_i^{(0)}(\mathcal{R}) &= \frac{K_i(\hat{\mathcal{R}})}{2\pi}\,\omega_i^{(1)}(\mathcal{R}) 
\end{align}
in the general form required for an anisotropic shape \cite{HansenGoosM2009}. 
A point on the boundary $\partial\mathcal{B}_i$ of body $\mathcal{B}_i$ with orientation $\varphi$ in the direction of the unit vector $\hat{\bvec{r}}=\bvec{r}/|\bvec{r}|$ is denoted by $\bvec{R}_i(\hat{\mathcal{R}})$, with $\hat{\mathcal{R}}$ short for $(\hat{\bvec{r}}, \varphi)$.
At this point, $K_i(\hat{\mathcal{R}})$ is the curvature and $\bvec{n}_i(\hat{\mathcal{R}})$ is the vector normal to the boundary. The orientation-dependence of the weight functions $\omega_i^{(\nu)}({\mathcal{R}})$ can be treated as described for three dimensions~\cite{WittmannM2014,Wittmann2015} or by considering each discrete orientation as an individual species.

Here we briefly outline the idea behind FMMT~\cite{WittmannMM2015,WittmannMM2015b}. For a more detailed description of the mathematical background see Refs.~\onlinecite{SchneiderW2008,Wittmann2015}. First we  identify in any spatial dimension the Mayer function $-f_{ij}=\chi(\mathcal{I}_{ij})$ with the Euler characteristic $\chi(\mathcal{I}_{ij})=\int\varPhi_0(\mathcal{I}_{ij},\upd\bvec{r})$ of the intersection. The latter can be further written as the spatial integral of the local curvature measure $\varPhi_0$, which is closely related to the weight function $\omega_i^{(0)}$ when evaluated for a body $\mathcal{B}_i$~\cite{WittmannMM2015b}. Applying in two dimensions the translative integral formula (iii) to Eq.~\eqref{eq_ldlimit} results for any orientations $\varphi_1$ and $\varphi_2$ in the decomposition
\begin{align}
 \label{eq_fijMM0}
 &-\iint f_{ij}(\mathcal{R}_1, \mathcal{R}_2)\, \upd\bvec{r}_1\, \upd\bvec{r}_2 \\\nonumber &=\iiint\upd\bvec{r} \sum_{k=0}^2  \varPhi^{(0)}_{k,2-k}(\bar{\mathcal{B}}_i(\bvec{r},\varphi_1), \bar{\mathcal{B}}_j(\bvec{r},\varphi_2);\upd (\bvec{r}_1,\bvec{r}_2))\,,
\end{align}
 defining an inverted body as $\bar{\mathcal{B}}_i(\bvec{r},\varphi_1):=2\bvec{r}-\mathcal{B}_i(\bvec{r},\varphi_1)$ and introducing the mixed measures $\varPhi^{(0)}_{k,2-k}$ \cite{WittmannMM2015,WittmannMM2015b}. For the precise definitions of $\varPhi_0$ and $\varPhi^{(0)}_{k,2-k}$, see Ref.~\onlinecite{SchneiderW2008}.

It can be shown \cite{WittmannMM2015b} that for $k=0$ and $k=2$ the expression on the right-hand side of Eq.~\eqref{eq_fijMM0} factorizes into a convolution product 
\begin{equation}\label{eq_conv}
 \omega_i^{(\nu)} \otimes \omega_j^{(\mu)}=\int\!\upd\bvec{r}'\,\omega_i^{(\nu)}(\bvec{r}'-\mathcal{R}_1)\;\omega_j^{(\mu)}(\bvec{r}'-\mathcal{R}_2)
\end{equation}
(integrated over $\upd \bvec{r}_1$ and $\upd \bvec{r}_2$) of the scalar weight functions with labels 0 and 2, where $(\bvec{r}-\mathcal{R}_1)$ is short for $(\bvec{r}-\bvec{r}_1,\varphi_1)$. In a similar way, we can define from $\varPhi^{(0)}_{1,1}$ the mixed weight function \cite{WittmannMM2015b}
\begin{equation} 
 \begin{split}
 &\varOmega^{(11)}_{ij}(\mathcal{R}_1,\mathcal{R}_2)\\
 &=\frac{\arccos(\bvec{n}_i\cdot\bvec{n}_j)}{2\pi}\left|\bvec{n}_i\times\bvec{n}_j\right|\,\omega^{(1)}_i(\mathcal{R}_1)\,\omega^{(1)}_j(\mathcal{R}_2)\,,
 \label{eq_Mwf2}
 \end{split}
\end{equation}
where the vector product of the normals $\bvec{n}_i = \bvec{n}_i(\hat{\mathcal{R}}_1)$ and $\bvec{n}_j = \bvec{n}_j(\hat{\mathcal{R}}_2)$ can be calculated by adding a $z$ component equal to zero or according to $\left|\bvec{n}_i\times\bvec{n}_j\right|=\sin(\arccos(\bvec{n}_i\cdot\bvec{n}_j))$.
Thus we find the decomposition \cite{WittmannMM2015,WittmannMM2015b}
\begin{align}
 \label{eq_fijMM}
 -f_{ij} =\omega_i^{(0)} \otimes \omega_j^{(2)}+\omega_i^{(2)} \otimes \omega_j^{(0)} + \varOmega^{(1\otimes1)}_{ij}
\end{align}
of the Mayer function, where we define
\begin{equation}
 \varOmega^{(1\otimes 1)}_{ij}=\int\!\upd\bvec{r}'\,\varOmega^{(11)}_{ij}(\bvec{r}'-\mathcal{R}_1,\bvec{r}'-\mathcal{R}_2)
 \label{eq_effconv} 
\end{equation}
according to Eq.~\eqref{eq_conv} for two one-body weights.

To show that the same decomposition can be obtained from the Gauss-Bonnet theorem (ii), we recall the result
\begin{align} 
 \label{eq_decompfijED0a} 
 -2\pi f_{ij}
 =  \!\!\!\!\int\limits_{\partial\mathcal{B}_i\cap\mathcal{B}_j} \!\!\!\!\mathrm{d} l_i\,K_i 
 + \!\!\!\!\int\limits_{\mathcal{B}_i\cap\partial\mathcal{B}_j}\!\!\!\! \mathrm{d} l_j\,K_j
 + \!\!\sum_{\partial\mathcal{B}_i\cap\partial\mathcal{B}_j}\!\!\!\! \phi
\end{align}
of Ref.~\onlinecite{RothMO2012}, where $\phi=\arccos(\bvec{n}_i\cdot\bvec{n}_j)$ is the angle between the normal vectors at each intersection points and $\bvec{n}_{i/j}=\bvec{n}_{i/j}(\bvec{r}'-\mathcal{R}_{1/2})$. The only difference is that we here consider an arbitrary convex body rather than a HD.
This generalization does not violate the underlying assumption $-f_{ij}=\chi(\mathcal{I}_{ij})$.

The line integrals in Eq.~\eqref{eq_decompfijED0a} involving the curvature $K_i$ at the boundary of the intersection can be deconvoluted in the standard way of FMT~\cite{RothMO2012}. To see that the last term is equal to $\varOmega^{(1\otimes 1)}_{ij}$, we rewrite the sum as a pseudo three-dimensional line integral~\cite{HansenGoosM2010}
\begin{align} 
 \int \frac{\phi\sin\phi\,\upd s}{|\bvec{n}_i\times\bvec{n}_j|}= \int\!\upd\bvec{r}'\,\phi\sin\phi\,
 \omega_i^{(1)}(\bvec{r}'-\mathcal{R}_1)\,\omega_j^{(1)}(\bvec{r}'-\mathcal{R}_2)\,,
\end{align}
where $\sin\phi=|\bvec{n}_i\times\bvec{n}_j|$. Thus we have shown that Eq.~\eqref{eq_decompfijED0a} also results in the decomposition given by Eq.~\eqref{eq_fijMM}.

The manner in which the Mayer function is rewritten in  Eq.~\eqref{eq_decompfijED0a} depends on the dimensionality, as the Gauss-Bonnet theorem only applies to two-dimensional manifolds. 
In three dimensions, we set $-f_{ij}=\chi(\partial\mathcal{I}_{ij})/2$ for the two-dimensional boundary $\partial\mathcal{I}_{ij}$ of the three dimensional intersection \cite{HansenGoosM2009}. 
There is no obvious analog of strategy (ii) in other dimensions.
In contrast, FMMT (iii) provides a formal decomposition of the Mayer function in an arbitrary dimension \cite{WittmannMM2015,WittmannMM2015b}.
Finally, we note that, although the last term in Eq.~\eqref{eq_fijMM} still depends on two bodies simultaneously, 
the presented decomposition considerably facilitates the numerical implementation compared to the bare Mayer function. 
This is because the two-body weight functions exclusively depend on geometrical quantities of the single bodies, 
which, however, cannot be further simplified by factorization~\cite{WittmannMM2015,WittmannMM2015b} without considering an approximate expansion~\cite{RothMO2012,HansenGoosM2009,HansenGoosM2010}.

\subsection{Excess free energy \label{sec_phiex}}

Following the standard procedure in FMT, we define the weighted densities \cite{HansenGoosM2009,Rosenfeld1989}
\begin{equation}
 \label{eq_weighdenorient}
  n_{\nu}(\bvec{r}) = \sum_{i=1}^{\kappa} \int\!\upd\mathcal{R}_1 \,  \rho_i(\mathcal{R}_1)\: \omega_i^{(\nu)}(\bvec{r}-\mathcal{R}_1) \,,
\end{equation}
for the scalar weight functions in Eq.~\eqref{eq_wfscal} and the mixed weighted density \cite{WittmannMM2015,WittmannMM2015b}
\begin{align} 
 \label{eq_MWDgen}
 &N(\bvec{r}) =\\
 & \sum_{i,j=1}^{\kappa} \iint\!\upd\mathcal{R}_1\, \upd\mathcal{R}_2 \, \rho_i(\mathcal{R}_1)\, \rho_j(\mathcal{R}_2)\,\varOmega_{ij}^{(11)}(\bvec{r}-\mathcal{R}_1,\bvec{r}-\mathcal{R}_2) \nonumber
\end{align}
corresponding to Eq.~\eqref{eq_Mwf2}.
With the decomposition of the Mayer function from Eq.~(\ref{eq_fijMM}), we obtain the excess free energy density
$\Phi_{\text{exc}}=n_0n_3+\frac{1}{2}N$
in the low-density limit, Eq.~(\ref{eq_ldlimit}),
i.e., the FMT version of the Onsager functional in two dimensions.
Following the procedure for the HD functional~\cite{Rosenfeld1990,RothMO2012}, the extrapolation to higher densities results in
\begin{equation}
 \Phi_{\text{exc}}
 =-n_0\ln(1-n_2)+\frac{N}{2(1-n_2)}\,.
 \label{eq_PhiRF}
\end{equation}
The uncommon choice of the prefactor in the second term stems from the definition of the mixed weight function
according to the decomposition in Eq.~\eqref{eq_fijMM0} in terms of mixed measures.

Although we have the means to perform a free minimization of the functional in Eq.~\eqref{eq_PhiRF}, an expansion in terms of tensorial one-body weighted densities as in three dimensions~\cite{HansenGoosM2009} might prove fruitful. Such an approximation can be obtained in a completely analog way as for HD~\cite{RothMO2012}, as the structure of the decomposition in Eq.~\eqref{eq_decompfijED0a} is exactly the same. Thus we Taylor expand the term $\arccos(\bvec{n}_i\cdot\bvec{n}_j)\sin(\arccos(\bvec{n}_i\cdot\bvec{n}_j))/(2\pi)$ in Eq.~\eqref{eq_Mwf2} up to quadratic order in $\bvec{n}_i\bvec{n}_j$ and identify the vectorial
\begin{equation}
 \overrightarrow{\omega}_i^{(1)}(\bvec{r}) = \bvec{n}_i(\hat{\bvec{r}})\, \omega_i^{(1)}(\bvec{r})
\end{equation}
and tensorial 
\begin{equation}
 \overleftrightarrow{\omega}_i^{(1)}(\bvec{r}) = \bvec{n}_i(\hat{\bvec{r}})\bvec{n}^\text{T}_i(\hat{\bvec{r}})\, \omega_i^{(1)}(\bvec{r})
\end{equation}
weight functions to factorize each term of this expansion. The corresponding one-body weighted densities $\overrightarrow{n}_1$ and $\overleftrightarrow{n}_1$ 
are then calculated according to Eq.~\eqref{eq_weighdenorient}. Note that it is important to write here $\bvec{n}_i$ instead of $\hat{\bvec{r}}$, which is only equivalent to $\bvec{n}_i(\hat{\bvec{r}})$ for HD parametrized in polar coordinates.

Following Ref.~\onlinecite{RothMO2012}, we will consider the expansion coefficients as free parameters, which we adapt  for a one-component system to ensure
(I) the correct second virial coefficient of the homogeneous and isotropic fluid,
(II) the correct dimensional crossover to one dimension and
(IIIa) the best fit to the Mayer function of HD.
Conditions (I) and (II) do not depend on the specific shape, so we find
\begin{equation}
 N\approx\frac{2+a}{6\pi}n_1n_1+\frac{a-4}{6\pi}\overrightarrow{n}_1\cdot\overrightarrow{n}_1
 +\frac{2-2a}{6\pi}\mbox{Tr}\left[\overleftrightarrow{n}_1\overleftrightarrow{n}_1\right]\,,
 \label{eq_PhiRFapp}
\end{equation}
in agreement with the approximation for HD~\cite{RothMO2012}, where (IIIa) results in $a_\text{HD}=11/4$.

For general convex bodies, we should demand a weaker criterion than (IIIa), as the excluded area of two bodies depending on their intermolecular angle is not exactly represented, which corresponds to the Mayer function integrated over the particle positions. Hence, we will determine the final, shape-dependent, parameter $a$, as in three dimensions~\cite{HansenGoosM2010}, by requiring (IIIb) a minimal quadratic deviation from the exact excluded area. This criterion can be refined in various ways following the examples~\cite{HansenGoosM2010,WittmannMM2014,WittmannMM2015,WittmannMM2016} in three dimensions.
In order to further improve the general functional according to criteria (IIIa) and (IIIb), it becomes necessary to introduce additional parameters by including higher-order terms of the expansion of the mixed weight function, which we will not consider here.

\section{Numerical Methods}
\label{sec:methods}

In the following, we study the phase behavior of (one component, $\kappa=1$) hard discorectangles of aspect ratio $l=L/D$ in two dimensions, i.e., capped rectangles of length $L$ and width $D$ equal to the diameter of the capping disks. The  goal of our work is twofold.
Firstly, we demonstrate for the first time a free numerical minimization of a FMMT functional, which is exact in the low-density limit, including the transitions between spatially inhomogeneous phases of anisotropic hard particles. Secondly, we extend the available reference data \cite{BatesF2000} for a system of hard discorectangles by performing new detailed MC simulations, which resolve between smectic and crystalline phase at high density. The used numerical techniques are described below.

\begin{figure}
 \includegraphics[width=0.9 \linewidth]{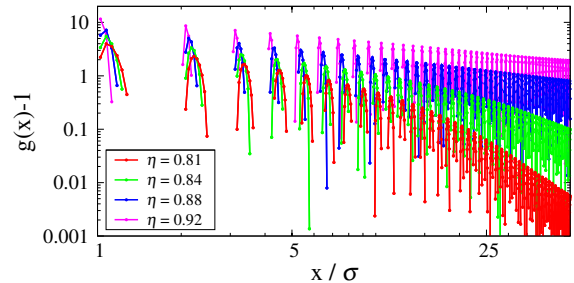}
 \caption{Translational ordering in the high-density smectic and crystalline phases for aspect ratio $l = 4$. At lower packing fractions $\eta \lesssim 0.85$ we observe a fast (exponential) decay of the in-plane pair correlation function $g_\parallel(r) - 1$, while for higher $\eta$ we find an algebraic decay.}
 \label{fig:orderplot}
\end{figure}

\subsection{Monte Carlo simulations}

We perform MC simulations of perfectly hard rods in the isobaric-isothermal ensemble, i.e., at constant pressure $p$, number of particles $N$, and temperature $T$. Each simulation contains $N = 5760$ particles in a rectangular simulation box with variable box lengths, and simulations were run for approximately $10^6$ MC sweeps (consisting of a rotation and translation move per particle, as well as a several volume moves). All simulations were started using a perfect crystalline lattice as the initial configuration. Overlaps were detected using the two-dimensional equivalent of the algorithm introduced by Vega {\it et al.} for spherocylinders \cite{VegaL1994}.

In the simulations, we measure the pair correlation function $g_\parallel(r)$ along a crystalline or smectic layer, averaged over the width of a single layer. We then plot $g_\parallel(r) - 1$ as a function of the distance $r$, and investigate how the oscillations decay towards zero at large distances. In particular, we associate exponential decay (indicating short-ranged positional order) with the smectic phase, and algebraic decay (associated with quasi-long-range order) with the crystalline phase. Figure \ref{fig:orderplot} shows typical examples of correlation functions for aspect ratio $l = 4$. For all aspect ratios $l \ge 2$, we observe a crossover from exponential to algebraic decay in the correlation functions as the packing fraction increases. We estimate the transition line between the smectic and crystalline phases by extracting for each aspect ratio the packing fraction at which this crossover occurs. 

In our simulations, we observe noticeable diffusion of particles between layers in the smectic phase, but essentially no diffusion in the crystal phase. This suggests that our observation of a transition to a crystalline phase might occur simply when our simulation are too short to sample the transfer of particles between layers. This would ensure that the number of particles per layer in our simulation is artificially fixed, favoring a crystalline state. To ensure that this effect does not meaningfully affect our result, we repeated simulations for several aspect ratios using shifted periodic boundary conditions, which facilitate transfer of particles between layers. Our results show no significant differences in the transition density measured using the two different approaches.

\subsection{Density functional theory}

Using the full expression \eqref{eq_MWDgen} for $N(\bvec{r})$ in the excess free-energy density \eqref{eq_PhiRF}, we minimize the grand-canonical free-energy functional in real space with respect to $\rho(\mathcal R)$ by analogy with Ref.~\onlinecite{SittaSWL2016} using the following Picard iteration scheme \cite{Roth2010}:
\begin{equation}
 \begin{split}%
 \rho^{(i+1)}(\mathcal R) &= (1-\tilde{\alpha}) \rho^{(i)}(\mathcal R) \\
 &\quad\, + \tilde{\alpha} \frac{1}{\Lambda^2}\exp{\!\Big(\beta \mu^{(i)} - \frac{\delta \beta \mathcal{F_{\mathrm{exc}}}}{\delta \rho(\mathcal R)}\Big)}
 \end{split}%
\end{equation}
with the mixing parameter $\tilde{\alpha} \le 0.01$, $\Lambda$ set to $D$, and the functional derivative $\frac{\delta \beta \mathcal{F_{\mathrm{exc}}}}{\delta \rho(\mathcal R)}$ (see also Eq.~(58) in Ref.~\onlinecite{WittmannMM2015b}). The chemical potential $\mu^{(i)}$ is recalculated in every iteration step to maintain the desired area fraction and converges to a finite value in the iteration. As in previous works \cite{OettelDBNS2012,HaertelOREHL2012, SittaSWL2016}, we combine this iteration with a direct inversion in the iterative subspace (DIIS) \cite{Ng1974,Pulay1980,Pulay1982,KovalenkoTNH1999} to improve the convergence.
The resolution of the spatial grid was chosen as $\Delta x = \Delta y \approx 0.03 D$ and the discrete orientations of the particles are chosen in equidistant steps of $\Delta \phi = 2 \pi/48$.

\section{Results for the phase diagram of hard discorectangles}
\label{sec:results}

The functional, Eq.~\eqref{eq_PhiRF}, based on the expansion in Eq.~\eqref{eq_PhiRFapp}, can be minimized analytically for hard discorectangles when we assume a homogeneous density.
Demanding that condition (IIIb) from Sec.~\ref{sec_phiex} holds, the remaining parameter becomes $a=3$ for any aspect ratio $l$ of the discorectangles. Note that in the HD limit, $l\rightarrow0$, where there is no distinction between an isotropic and nematic phase, condition (IIIb) is equivalent to (I), so that the parameter $a=a_\text{HD}=11/4$ can be used to fulfill (IIIa) instead~\cite{RothMO2012}. 
However, the present choice $a=3$ was found in Ref.~\onlinecite{RothMO2012} to be even more consistent with the simulation data for the bulk pressure of the HD crystal.
Moreover, the excluded area of parallel discorectangles can be exactly represented by choosing $a=4$.

By analogy with functional in three dimensions, we expect that the choice $a=3$ will provide reliable results for the isotropic and moderately ordered nematic phases but will not allow us to describe a stable smectic phase~\cite{WittmannMM2014,WittmannMM2016}. The latter is only possible qualitatively for $a=4$, ensuring that the free energy per particle does not diverge in the limit $l\rightarrow\infty$. However, this parameter will result in a poor description of the homogeneous phases~\cite{WittmannMM2014,WittmannMM2016}, which is most apparent by comparing to $a_\text{HD}$ in the HD limit. 

To avoid the ambiguity of choosing a proper value of $a$  our main objective is perform a free numerical minimization of the full functional from Eq.~\eqref{eq_PhiRF} with the mixed weighted density from Eq.~\eqref{eq_MWDgen}, which is feasible in two dimensions.  The employed algorithm is described in Sec.~\ref{sec:methods}. As a first step, however, we will demonstrate the utility of expanding the functional by calculating a closed expression for the isotropic--nematic transition line.

To characterize the homogeneous phases, we represent the density $\rho(\varphi)=\rho \, g(\cos\varphi)$ of in terms of a normalized orientational distribution function $g(\cos\varphi)$.  
The two-dimensional nematic order parameter is conveniently defined as
\begin{equation} 
 S=\frac{2}{\pi}\int_0^{\pi/2} \!\!\!\!\! \upd\varphi \left(2\cos^2\varphi-1\right)\,g(\cos\varphi)\,.
 \label{eq_Snem}
\end{equation}
For discorectangles we obtain the weighted densities
\begin{align}
 n_2&=\rho \left(LD+\frac{\pi}{4}D^2 \right)
 = \eta\, , \cr
 n_1 &= \rho\,(2L+\pi D)\, ,\ \ \ n_0= \rho \, , \cr
(\overleftrightarrow{n}_1)_{11}&=\rho\left(L(1+S)+\frac{\pi}{2}D\right)\, ,\cr
 (\overleftrightarrow{n}_1)_{22} & = \rho\left(L(1-S)+\frac{\pi}{2}D\right)\, 
 \label{eq_gewdich}
\end{align}
where $\eta$ denotes the packing fraction. For a given aspect ratio $l$, the (nematic) free energy thus becomes a function of $\eta$ and $S$ when we use the approximation in Eq.~\eqref{eq_PhiRFapp}.

Minimization with respect to the orientational distribution function \cite{HansenGoosM2009} results in
\begin{equation}  
 g(\alpha,\cos\varphi)=\frac{\exp\left(\alpha^2(2\cos^2\varphi-1)\right)}{I_0(\alpha^2)}  \,,
 \label{eq_gnem}
\end{equation}  
where $I_n$ denotes the modified Bessel function of the first kind, which follows from the normalization condition $\int_0^{\pi/2}\upd\varphi \,g(\cos\varphi)=\pi/2$. The parameter $\alpha(\eta,l)$ then follows from the self-consistency equation
\begin{equation}  
 \alpha^2:=-\frac{\partial\Phi_\text{ex}(\eta,S,l)}{\rho\,\partial S}\,.
 \label{eq_alphanem}
\end{equation} 
Inserting Eq.~\eqref{eq_gnem} into Eq.~\eqref{eq_Snem} we obtain the nematic order parameter
\begin{equation} 
 S(\alpha)=\frac{I_1(\alpha^2)}{I_0(\alpha^2)}=\frac{1}{2}\alpha^2-\frac{1}{16}\alpha^6+\mathcal{O}(\alpha^{10})
 \label{eq_SofALPHA}
\end{equation}
as a function of $\alpha$.

\subsection{Isotropic--nematic transition \label{sec_IN}}

In order to study the isotropic--nematic transition, one has to solve Eq.~\eqref{eq_alphanem}.
For the functional from Eqs.~\eqref{eq_PhiRF} and~\eqref{eq_PhiRFapp}, there is at most one (stable) solution to Eq.~\eqref{eq_alphanem} at a given density, which is not the case in three dimensions. This can be easily seen  by rewriting the condition in the generic form $\alpha^2-CS(\alpha)=0$ with a positive parameter $C\simeq C(\eta,l)$. The position $C_\text{min}(\alpha)$ at which the expression on the left-hand-side becomes minimal increases monotonously with increasing $\alpha$.
Therefore, at $\alpha=0$ the isotropic and nematic solutions are indistinguishable, denoting a second-order phase transition, as it is expected from computer simulations~\cite{FrenkelE1985,BatesF2000}, although also first-order transitions between the isotropic and nematic phase are discussed in the literature~\cite{Vink2007, WensinkV2007, FishV2010}.
Nevertheless, in three dimensions, the corresponding result for $S(\alpha)$ admits a non-monotonic behavior of $C_\text{min}(\alpha)$, indicating that the nematic phase is only metastable at small order parameters, i.e., the isotropic--nematic transition is of first order \cite{HansenGoosM2009}.

Solving Eq.~\eqref{eq_alphanem} for $\eta$ yields a closed expression for the packing fraction
\begin{equation} 
 \eta_\text{N}(\alpha)=
 \left(1+\frac{8l^2(a-1)I_1(\alpha^2)}{3\pi(4l+\pi) \alpha^2I_0(\alpha^2)}\right)^{-1}
 \label{eq_etaN}
\end{equation}
at which the nematic phase is stable for a given $\alpha$.
Numerically inverting $\eta_\text{N}(\alpha)$ and comparing to Eq.~\eqref{eq_SofALPHA}, we can calculate the nematic order parameter $S(\eta)$ as a function of the packing fraction. In the limit of vanishing orientational order, we obtain the packing fraction
\begin{equation} 
 \eta_\text{IN}:=\lim\limits_{\alpha\rightarrow0}\eta_\text{N}(\alpha)=
 \left(1+\frac{4l^2(a-1)}{3\pi(4l+\pi)}\right)^{-1}
 \label{eq_etaIN}
\end{equation}
at the second-order isotropic--nematic transition of hard discorectangles in two dimensions for an arbitrary aspect ratio $l$ and the parameter $a$. Obviously, with increasing the aspect ratio the transition density decreases down to the scaled density
\begin{equation} 
 c_\text{N}:=\lim\limits_{l\rightarrow\infty}\eta_\text{IN}l=\frac{3\pi}{a-1}\,.
 \label{eq_cIN}
\end{equation}
obtained in the Onsager limit $l\rightarrow\infty$.

As the isotropic--nematic transition in two dimensions is of second order, the influence of higher-order terms in the expansion of the mixed weighted density from Eq.~\eqref{eq_PhiRFapp} is negligible, if the appropriate value $a=3$ is chosen in a way that it ensures that the leading term in the order-parameter dependence is retained. Therefore, the result for $\eta_\text{IN}$ given by Eq.~\eqref{eq_etaIN} with $a=3$ is equivalent to that obtained with the full functional from Eq.~\eqref{eq_PhiRFapp} based on the exact two-body representation.
This is nicely confirmed for infinitely long rods where the Onsager result~\cite{KayserR1978} for the transition density is given by Eq.~\eqref{eq_cIN} with $a=3$.
A more detailed explanation has been given for the analogous three-dimensional case~\cite{WittmannMM2015b,Wittmann2015}, addressing the limit of metastability of the isotropic phase. Up to the first non-vanishing term in the parameter $\alpha$ we can write
\begin{equation} 
 \eta_\text{N}(\alpha)-\eta_\text{IN}=\frac{6l^2\pi(4l+\pi)(a-1)}{(4l^2(a-1)+3\pi(4l+\pi))^2} S^2 +\mathcal{O}(\alpha^8)
 \label{eq_DetaIN}
\end{equation}
with the help of Eq.~\eqref{eq_SofALPHA}. This result suggests that the nematic order parameter $S$ approaches the critical point with a critical exponent of $1/2$.

 \begin{figure}[t]
 \includegraphics[width=0.45\textwidth]{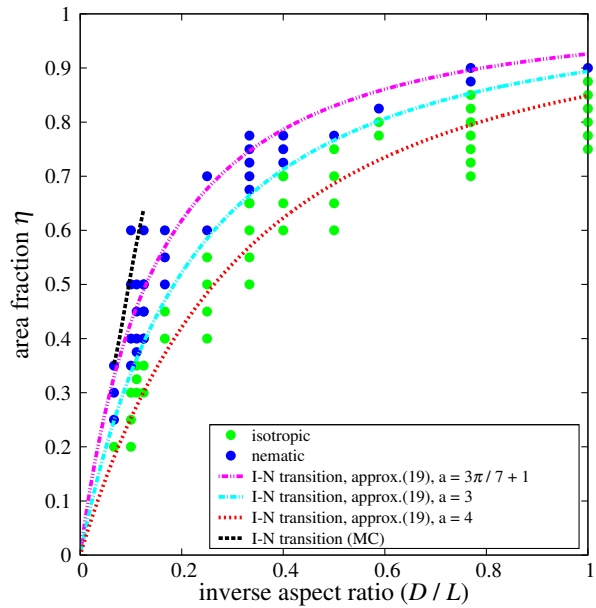}
 \caption{Packing fraction at the second-order isotropic--nematic transition of two-dimensional hard discorectangles from the analytic prediction given by Eq.~\eqref{eq_etaIN} (lines) for the parameters $a=3\pi/7 + 1\approx2.35$ (obtained  to fit Eq.~\eqref{eq_cIN} to the simulation result~\cite{FrenkelE1985} in  the Onsager limit, magenta), $a=3$ (cyan), and $a=4$ (red) as a function of the inverse aspect ratio $l^{-1}=D/L$. The approximation \eqref{eq_PhiRFapp} with $a=3$ matches the isotropic--nematic transition of the full functional with Eq.~\eqref{eq_MWDgen} (points), which was numerically evaluated under the constraint of a spatially homogeneous density. Data from MC simulations \cite{BatesF2000} are shown for comparison (black line).}
 \label{fig_IN}
\end{figure} 
 
In Fig.~\ref{fig_IN} we show the density at the isotropic--nematic transition of hard discorectangles given by Eq.~\eqref{eq_etaIN} as a function of the (inverse) aspect ratio $D/L$ for different values of the parameter $a$. 
Our full minimization confirms the analytical finding that the transition is of second order. As discussed above, the results agree perfectly with those for $a=3$ within the numeric error. For instance, at $l=9$ we find $\eta_\text{IN}=0.363$ and numerically $\eta_\text{IN}=0.365\pm0.002$.
However, compared to the simulation data~\cite{FrenkelE1985,BatesF2000} the DFT predicts much smaller values.
This discrepancy is comparable to the inaccuracy of the Onsager functional~\cite{KayserR1978} due to disregarding the virial coefficients higher that the second, which do not vanish in two dimensions.
From this perspective, Eq.~\eqref{eq_cIN} provides a simple method to fix the parameter $a$ to recover simulation result $c_\text{IN}\approx 7$ for $l\rightarrow\infty$~\cite{FrenkelE1985}, which appears more reasonable than fitting to simulation data at finite aspect ratio, as proposed in three dimensions~\cite{HansenGoosM2009}, but is still empirical in nature. Indeed, accordingly choosing $a=3\pi/7+1$ results in a better agreement with the simulation data for  rods of finite thickness, but deviates more and more with decreasing aspect ratio. Since this purely empirical approach is also inconsistent with the proper implementation of FMMT we will not further discuss it here.  On the other hand, choosing $a=4$ results in the poorest functional for the isotropic--nematic transition.

\begin{figure}[]
\includegraphics[width=0.45\textwidth]{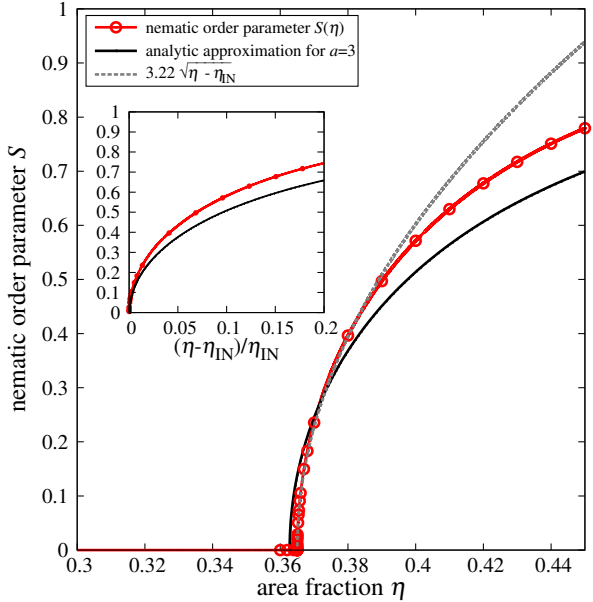}
\caption{Nematic order parameter $S(\eta)$ for discorectangles with aspect ratio $l = 9$ close to the area fraction of the isotropic--nematic transition $\eta_t$ as a function of the area fraction $\eta$. Both numeric results for Eq.~\eqref{eq_MWDgen} (red) and analytic results for the approximation \eqref{eq_PhiRFapp} with $a=3$ (black) show a second order transition between the isotropic and the nematic phase. Close to this transition, the nematic order parameter can be fitted with a squareroot function (gray), showing a critical exponent of $1/2$, which agrees with that suggested by Eq.~\eqref{eq_DetaIN}. The inset compares the shape of $S(\eta)$ for Eq.~\eqref{eq_MWDgen} and \eqref{eq_PhiRFapp} when rescaled.}
\label{fig:nematic_order}
\end{figure} 
 
Now we study the nematic phase of discorectangles with aspect ratio $l = 9$ in more detail.
For various densities close to the isotropic--nematic transition, we compare in Fig.~\ref{fig:nematic_order} the nematic order parameter $S$ obtained according to Eq.~\eqref{eq_Snem} from a minimization of the full functional (red) and the analytical approximation with $a=3$ using Eqs.~\eqref{eq_SofALPHA} and~\eqref{eq_etaN} (black). We observe that beyond the common (up to a horizontal shift due to the numeric error) transition point with $S(\eta_\text{IN})=0$ the numerical result for the order parameter increases faster than that of the approximation in terms of rank-two tensors.
Fitting $b \sqrt{\eta - \eta_\text{IN}}$ with fit parameter $b\approx3.22$ (gray) to the numeric data shows a very good agreement close to $\eta_\text{IN}$ and also points to a critical exponent of $1/2$. From Eq.~\eqref{eq_DetaIN} we find $\tilde{b}\approx2.94<b$ for the analytical approximation.
  
The reason for the stronger increase of the order parameter in the numerical data is that the orientational distribution, Eq.~\eqref{eq_gnem}, and thus the expansion of the nematic order parameter $S$ in Eq.~\eqref{eq_SofALPHA} is only exact at leading order in the orientational anisotropy, i.e., up to the quadratic term in $\alpha$.
As demonstrated in three dimensions~\cite{WittmannMM2015b,Wittmann2015}, it is possible to include tensors of higher rank to the expansion from Eq.~\eqref{eq_PhiRFapp} in a systematic way (suitably chosen correction parameters). This results in the presence of additional order parameters and a more accurate analytic solution for the nematic orientational distribution. For example, the nematic order parameter in Fig.~\ref{fig:nematic_order} calculated from such an approach would converge to the data from a free minimization of the full functional.

\subsection{Inhomogeneous Phases}

\begin{figure*}[]
 \includegraphics[width=0.765\textwidth]{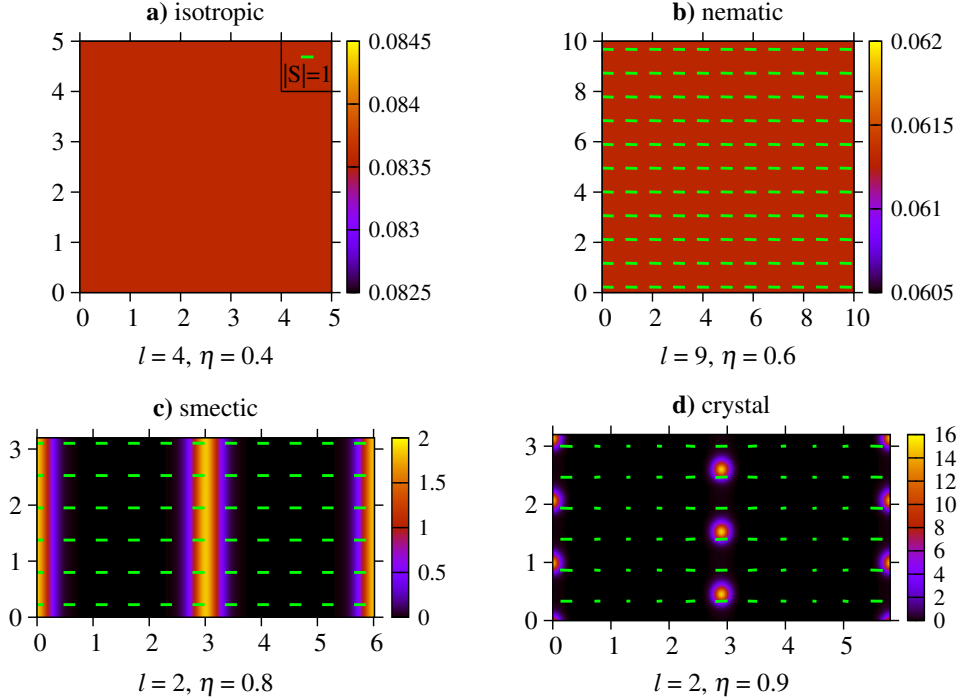}
 \caption{\label{fig:DFT_phases}Density and orientation profiles for various aspect ratios $l$ and area fractions $\eta$. The orientation integrated center-of-mass density is shown in the color plot, while the local mean orientation is indicated with green dashes for (a) the isotropic phase at $l=4$, $\eta=0.4$, (b) the nematic phase at $l=9$, $\eta=0.6$, (c) the smectic phase at $l=2$, $\eta=0.8$, (d) the crystalline phase at $l=2$, $\eta=0.9$. For scale, a dash with length corresponding to perfect nematic order ($|S|=1$) is drawn in (a).}
\end{figure*} 

\begin{figure}[t]
 \includegraphics[width=0.45\textwidth]{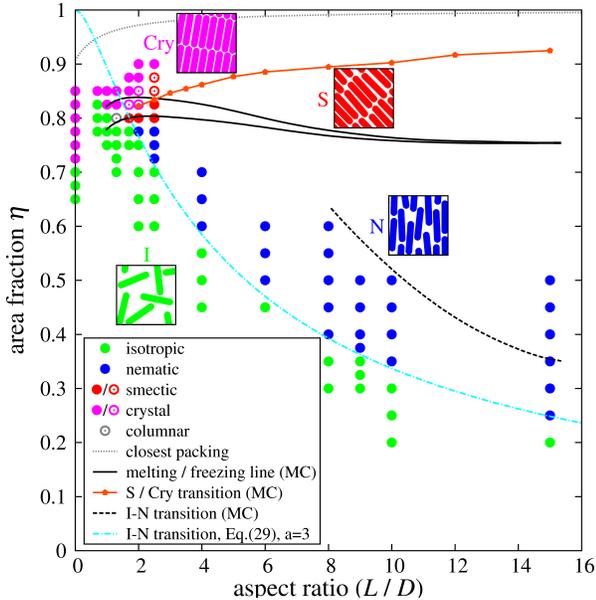}
 \caption{The phase diagram for discorectangles as a function of aspect ratio and area fraction is shown for FMMT data (points) and compared with MC data (lines). For the transition between a solid phase (S or Cry) and a fluid phase (I or N), the corresponding melting and freezing lines for MC (black solid lines, adopted from Ref.~\onlinecite{BatesF2000}) are in very good agreement with DFT. At sufficiently high aspect ratios, we find for the solid a transition from a smectic (red) to a crystalline (purple) phase for both MC (red line) and DFT. Close to this transition, the energy difference between the smectic and crystalline phase was close to our numeric error bars. Those points are displayed with open symbols for the slightly dominating phase. The stability of the nematic phase (blue) is overestimated when compared with MC data (black dashed line, adopted from Ref.~\onlinecite{BatesF2000}). Though the topology is identical for both DFT and MC, except that we find a columnar phase once in the DFT (gray). The closest packing (gray pointed line) is shown for comparison. Sketches for the phases are shown in the insets.
 }
 \label{fig:phase_diagram}
\end{figure}

Taking now also spatially inhomogeneous phases into consideration, it is no longer possible to obtain an accurate analytic solution of the functional. Instead, we perform a full numerical minimization of the functional including the full expression, Eq.~\eqref{eq_MWDgen}, for the mixed weighted density. We find in total four different phases for both DFT and MC: (a) an isotropic phase with neither orientational nor spatial order, (b) a nematic phase with orientational but no spatial order, (c) a smectic phase with orientational order and spatial order in one dimension, and (d) a crystalline phase with both orientational order and spatial order in two dimensions. Typical DFT profiles for these four phases are shown in Fig.~\ref{fig:DFT_phases} and particle resolved sketches for these phases are shown in the insets of Fig.~\ref{fig:phase_diagram}. The phase diagram for varying aspect ratios ($l=L/D$) and area fractions ($\eta$) is shown in Fig.~\ref{fig:phase_diagram}.

The isotropic phase (I, green) is dominating at low area fractions. For all aspect ratios at sufficiently high area fractions, the discorectangles freeze into a crystal (Cry, purple), with layers of discorectangles. Particles of adjacent layers are shifted by half a particle width, allowing the rounded caps of one particle to fill the voids between two rounded caps in each adjacent layer (see density profile in Fig.~\ref{fig:DFT_phases} (d), or the inset in Fig.~\ref{fig:phase_diagram}). Such a crystal allows for the closest packing (gray dotted line in Fig.~\ref{fig:phase_diagram}). The closest packing $\eta_\text{cp}$ as a function of the aspect ratio $l$ is given by
\begin{equation} 
 \eta_\text{cp} = \frac{l + \pi/4}{l + \sqrt{3}/2}\,.
 \label{eq_etaCP}
\end{equation} 
For large aspect ratios, a nematic phase (N, blue) is found for intermediate area fractions for both DFT and MC, although the stability of the nematic phase is overestimated by DFT when compared with MC data (black dashed lines Ref.~\onlinecite{BatesF2000}), as discussed in Sec.~\ref{sec_IN}. 

For the first-order phase transition between a solid phase (either smectic or crystal) and a fluid phase (either isotropic or nematic), MC data are taken from Bates and Frenkel \cite{BatesF2000}. These ``melting/freezing lines'' (black solid lines in Fig.~\ref{fig:phase_diagram}) are coexistence lines, which were calculated via free energy calculations for $l \le 7$ and extrapolated to larger aspect ratios \cite{BatesF2000}. For this phase transition, the agreement between the MC data and our DFT is very good. In particular, in the HD limit $(l=0)$, the transition density between the isotropic and the crystalline phase matches the values found in early
computer simulations \cite{AlderW1962} and DFT calculations based on Eq.~\eqref{eq_PhiRFapp}~\cite{RothMO2012}.

For small aspect ratios ($l \le 1$), the discorectangles directly freeze into a crystal when increasing the area fraction. At higher aspect ratios ($l \ge 1.7$), the discorectangles freeze first into a smectic phase (S, red) with orientational and translational order in just one spatial dimension, before they cross over to the crystal at even higher area fractions. We find this behavior for both DFT and our refined MC. Close to this transition, the energy difference between the smectic and crystalline phase is close to our numeric error bars. This uncertainty is depicted in Fig.~\ref{fig:phase_diagram} with open symbols indicating the slightly dominating phase. At $l=1.3$ (close to the isotropic--smectic--crystal triple point) we find in DFT a columnar phase, with particle alignment parallel to the density layer, instead of a smectic phase at a single state point between the isotropic and the crystal. We do not observe this in MC. 

Note that the MC data in Fig.~\ref{fig:phase_diagram} for the melting of the solid does not take into account the possibility of melting via a Kosterlitz-Thouless (KT) dislocation unbinding mechanism, which would not be visible in DFT~\cite{RothMO2012}, but could influence simulations. However, such a melting scenario is only likely to occur for very short rods, close to the HD limit where a KT crystal--hexatic and a first-order hexatic--isotropic transition is predicted \cite{BernardK2011, ThorneyworkAAD2017}. Reference~\onlinecite{BatesF2000} reported no evidence of topological defects even for $l = 1$, and hence in the regime explored with simulations here we do not expect this scenario. In contrast, the nematic--isotropic transition does occur via a continuous KT transition \cite{BatesF2000}.

\section{Conclusions \label{sec:conclusions}}

In conclusion, we have predicted the bulk phase diagram of two-dimensional hard rods from Fundamental Mixed Measure Theory and found a stable isotropic, nematic, smectic, and crystalline phase depending on the particle aspect ratio and density. In general, it is mandatory to use a free minimization technique to obtain the correct minimizing equilibrium state in the density functional theory. At intermediate area fractions, the second-order isotropic--nematic transition is equally described by an analytical curve found from a simple expansion of the functional. The density functional results for the phase diagram agree well with our MC calculations, which also show a stable smectic phase.

For the first time, we have implemented a free minimization of the two-dimensional version of FMMT, which is not feasible in higher dimensions.
In further contrast to the three-dimensional case, the mixed weighted density does not vanish in the HD limit~\cite{TarazonaR1997}, and thus our approach also provides the most accurate way to study the crystallization of the HD fluid within the framework of FMT. 
Moreover, the value of the free parameter in the computationally more efficient expanded functional from Ref.~\onlinecite{RothMO2012} derived here appears to give better results for the HD crystal~\cite{RothMO2012} and the liquid--crystal surface tension~\cite{Oettel_Lin}. 

For the inhomogeneous phases of very long rods at high densities, a free minimization of FMMT becomes more demanding.
Following the examples in three dimensions, this region of the phase diagram can be explored more efficiently by systematically expanding the mixed weighted density for intermediate aspect ratios~\cite{MarechalDD2017} 
or by a linearization in the orientation dependence for highly aligned systems of very long rods~\cite{WittmannMM2015,Wittmann2015}.
Another possible simplification would be to perform an (approximate) parametrized minimization, in combination with the decoupling approximation~\cite{WittmannMM2014}. 
The latter is well justified for the smectic phase, since Fig.~\ref{fig:DFT_phases}c indicates that the orientational order is independent of the spatial coordinate.
This procedure could allow for an analytic calculation in the limit of strong alignment~\cite{WittmannMM2015,Wittmann2015}.

We have seen that the onset of crystallization is predicted very accurately for the aspect ratios considered here, which should also be the case for longer rods since FMMT recovers the cell theory limit~\cite{WittmannMM2016}. 
 Note that, for some versions of the cubic term in the density, the three-dimensional FMMT can diverge in an unphysical way when applied to highly aligned long rods~\cite{WittmannMM2014,WittmannMM2016}.
 There is no such pitfall for FMMT (and the expanded form with $a=4$) in two dimensions.
By construction, FMMT also reduces to scaled particle theory for a homogeneous fluid, which is known to overestimate the pressure at finite density, and to Onsager theory for infinitely long rods, which omits all relevant virial coefficients higher than the second.
As a result, the nematic phase is severely overstabilized compared to the isotropic fluid and we expect that the nematic--smectic transition of long rods predicted by FMMT will occur at smaller densities than found in computer simulations, which is also the case in three dimensions~\cite{WittmannMM2015}.

In the future, the theory should be applied and generalized towards different situations: first of all, other particle shapes such as two-dimensional ellipses \cite{FoulaadvandY2013,BautistaCarbajalO2014} or  rectangles with sharp edges \cite{GonzalezPintoMRV2013,SittaSWL2016} can be considered, as well as mixtures between particles of different sizes \cite{Rosenfeld1990, DijkstravRE1999} or shapes \cite{vanderKrooijL2000}. Second, our bulk phase diagram provides the starting point for a microscopic theory of interfaces \cite{WittmannM2014,HaertelOREHL2012} between two coexisting phases which show interesting translational and orientational structures \cite{PraetoriusVWL2013}. Third, our density functional theory can be generalized towards dynamical density functional theory \cite{MarconiT1999, MarconiT2000, ArcherE2004, MarconiM2007, EspanolL2009, WittkowskiL2011, GoddardNSYK2013} describing translational and orientational Brownian dynamics of rods \cite{RexWL2007}.
Finally our results for the phase diagram can in principle be verified by experiments using layers of monodisperse sterically-stabilized colloidal rod-like particles \cite{LinCPSWLY2000,GalanisNLH2010,HermesVLVvODvB2011,ZvyagolskayaAB2011,BesselingHKdNDDIvB2015,MuellerdlHRH2015,WalshM2016}. Another macroscopic option are shaken granular rods \cite{CruzHidalgoZMP2010,HernandezNavarroIMST2012} on a substrate which resemble equilibrium phase behavior.

\acknowledgments
 We thank Axel Voigt, Martin Oettel and Shang-Chun Lin for helpful discussions.
Financial support from the German Research Foundation (DFG) is gratefully acknowledged within the Project LO 418/20-1.
Moreover, R.\ W.\ gratefully acknowledges funding provided by the Swiss National Science Foundation and fruitful discussions with Joseph Brader and Matthieu Marechal.

\bibliography{refs}

\end{document}